\documentclass[10pt]{emulateapj}
\usepackage{apjfonts}
\usepackage{graphicx}
\usepackage{psfig}

\newcommand{\begit}{\begin{itemize}}
\newcommand{\enit}{\end{itemize}}
\newcommand{\begen}{\begin{enumerate}}
\newcommand{\enen}{\end{enumerate}}
\newcommand{\beq}{\begin{equation}}
\newcommand{\eeq}{\end{equation}}
\newcommand{\beqa}{\begin{eqnarray}}
\newcommand{\eeqa}{\end{eqnarray}}

\begin{document}

\title{Assessing The Starburst Contribution to the $\gamma$-Ray \&
Neutrino Backgrounds}

\author{Todd A.~Thompson,\altaffilmark{1,2}
Eliot Quataert,\altaffilmark{3}
Eli Waxman,\altaffilmark{4} \& Abraham Loeb\altaffilmark{5}  }

\altaffiltext{1}{
Department of Astrophysical Sciences, Peyton Hall-Ivy Lane,
Princeton University, Princeton, NJ 08544; thomp@astro.princeton.edu}
\altaffiltext{2}{Lyman Spitzer Jr.~Fellow}
\altaffiltext{3}{Astronomy Department
\& Theoretical Astrophysics Center, 601 Campbell Hall,
The University of California, Berkeley, CA 94720;
eliot@astro.berkeley.edu}
\altaffiltext{4}{Physics Faculty, Weizmann Institute of Science, Rehovot
76100, Israel;
waxman@wicc.weizmann.ac.il}
\altaffiltext{5}{
Astronomy Department, Harvard University, 60 Garden Street, Cambridge, MA
02138; aloeb@cfa.harvard.edu}

\begin{abstract}

If cosmic ray protons interact with gas at roughly 
the mean density of the interstellar
medium in starburst galaxies, then pion decay in starbursts is
likely to contribute significantly to the diffuse extra-galactic
background in both $\gamma$-rays and high energy neutrinos.  We
describe the assumptions that lead to this conclusion and clarify the
difference between our estimates and those of Stecker (2006).
Detection of a single starburst by {\it GLAST} would confirm the
significant contribution of starburst galaxies to the 
extra-galactic neutrino and $\gamma$-ray backgrounds.

\end{abstract}


\section{Introduction \& Proton Calorimetry}
\label{section:introduction}

Inelastic proton-proton collisions between cosmic rays and ambient nuclei
produce charged and neutral pions, which subsequently decay into
electrons, positrons, neutrinos, and $\gamma$-rays.  Neutral pion
decay accounts for most of the $\gamma$-ray emission from the Milky Way
at GeV energies (e.g., Strong et al.~2004).  The contribution of pion
decay to high energy emission from other galaxies --- and thus also to
the extra-galactic neutrino and $\gamma$-ray backgrounds --- remains
uncertain, however, because of the current lack of strong
observational constraints.  For example, aside from the LMC (Sreekumar
et al.~1992), EGRET did not detect $\gamma$-ray emission from star
formation in any other galaxy.  In two recent papers (Loeb \& Waxman
2006; Thompson, Quataert, \& Waxman 2006; hereafter LW and TQW,
respectively), we have estimated that starburst galaxies contribute
significantly to the extra-galactic neutrino and $\gamma$-ray backgrounds,
with most of the contribution arising from starburst galaxies at $z
\sim 1$.  TQW also provide detailed predictions for the $\gamma$-ray
fluxes from nearby starbursts, several of which should be detectable
with {\it GLAST}.  Using what seem to be similar arguments, however,
Stecker (2006) presents an upper limit on the starburst contribution
to the neutrino and $\gamma$-ray backgrounds that is a factor of $\approx
5$ below the predictions of LW and TQW. Here we clarify the difference
between these predictions (see \S\ref{section:other}).

The timescale for cosmic ray protons to lose energy via $p$-$p$
collisions with gas of number density $n$ cm$^{-3}$ is
$\tau_{pp}\approx7\times10^7/n$\,yr.  The importance of pion losses
depends on the ratio of $\tau_{pp}$ to the cosmic-ray proton escape
timescale, $\tau_{\rm esc}$.  The escape timescale for cosmic rays is
quite uncertain, but in starbursts it is probably set by advection
in a galactic superwind, rather than diffusion, as in the Galaxy.  If
correct, the escape timescale from a starburst with a wind of velocity
$V\approx300$\,km\,s$^{-1}$ and a gas scale height of
$h\approx100$\,pc is of order $\tau_{\rm esc}\approx
h/V\approx3\times10^5$\,yr.  This implies that if cosmic rays interact
with gas of mean density $n\gtrsim$100\,cm$^{-3}$, then
$\tau_{pp}\lesssim\tau_{\rm esc}$ and we expect that essentially all
cosmic ray protons are converted into charged and neutral pions and
their secondaries before escaping. In this limit, the galaxy is said
to be a ``cosmic ray proton calorimeter.''  If escape is instead due to
diffusion rather than advection, then the density threshold required
for a galaxy to be a proton calorimeter is likely to be significantly lower.

In the Milky Way, few GeV cosmic-ray protons only lose $\approx 10\%$
of their energy to pion production before escaping the galaxy.  For
this reason, LW and TQW focused on the high energy emission from
luminous starbursts.  These galaxies have much higher gas densities
and likely dominate the contribution of star-forming galaxies to the
total $\gamma$-ray and neutrino backgrounds.  For reference, we note
that the local starbursts NGC\,253 and M\,82 have $n\gtrsim400$\,cm$^{-3}$,
and the nuclei of Arp\,220  have gas densities of order $10^4$\,cm$^{-3}$,
well in excess of that required by the above arguments for  $\tau_{pp}<\tau_{\rm esc}$.

Since the fraction of the proton energy supplied to secondary
electrons, positrons, neutrinos and $\gamma$-rays is determined by the
microphysics of the Standard Model, the $\gamma$-ray and neutrino fluxes
from starbursts depend primarily on astrophysical properties of star
forming galaxies via

\begin{enumerate}

\item{the fraction of a SN's canonical $10^{51}$ ergs of energy
supplied to cosmic ray protons, $\eta$,}

\item{the spectral index $p$ of the injected cosmic ray proton
distribution,}

\item{the evolution of the comoving star formation rate density of
the universe,
$\dot{\rho}_\star(z)$ (M$_\odot$ yr$^{-1}$ Mpc$^{-3}$),}

\item{and the fraction of all star formation that occurs in the proton
calorimeter limit as a function of cosmic time, $f(z)$.}

\end{enumerate}

For any $\eta$ and $p$, the $\gamma$-ray and neutrino emission from a
given proton calorimeter can be readily calculated, while the
cumulative extra-galactic backgrounds also depend on the star formation
history of the universe via $\dot{\rho}_\star(z)$ and $f(z)$.

As TQW and LW show, in the proton calorimeter limit the total
$\gamma$-ray and neutrino luminosities for an individual starburst can
be related to $L_{\rm TIR}$, the galaxy's total IR luminosity, by 
\beq
L_\gamma \approx(2/3) L_{\rm nu}\approx 1.5 \times 10^{-4}
\eta_{0.05}L_{\rm TIR},
\label{gamma_neutrino_fir}
\eeq
where $\eta_{0.05}=\eta/0.05$ (i.e., $5\times10^{49}\eta_{0.05}$\,ergs
per supernova is supplied to cosmic ray protons).  $\eta$ is
constrained by the requirement that the radio emission from secondary
electrons/positrons in starbursts be consistent with the observed
FIR-radio correlation (see, e.g., Yun et al.~2001).  For proton
calorimeters, $\eta \gtrsim 0.05$ is only possible if ionization,
bremsstrahlung, and/or inverse-Compton losses dominate synchrotron
losses for secondary electrons and positrons in starbursts (see TQW;
Thompson et al.~2006).

%%%%%%%%%%%%%%%%%%%%%%%%%%%%%%%%%%%%%%%%%%%%%%%%%%%%%%%%%%%%%%%%%%%%%%%%%%%%
\begin{figure}
\centerline{\hbox{\psfig{file=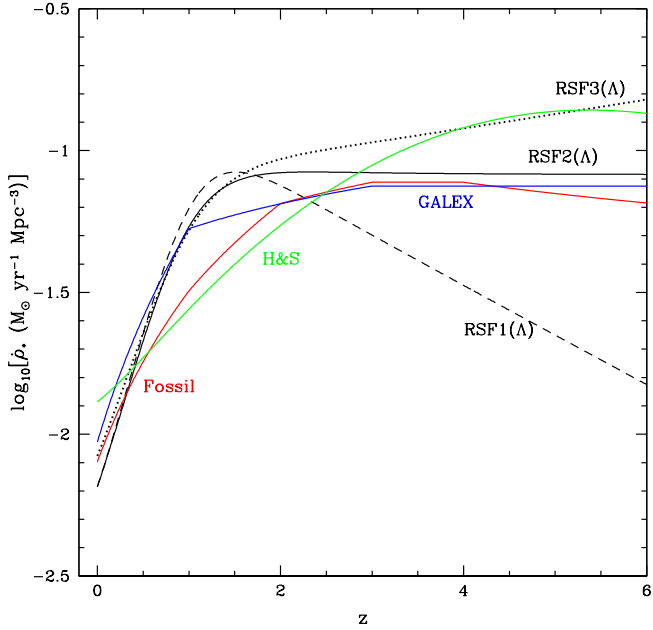,width=9cm}}}
\figcaption[x]{The comoving star formation rate density
$\dot{\rho}_\star(z)$ for a number of models and fits to existing
data: RSF1 ({\it dashed black}), RSF2 ({\it solid black}), and RSF3
({\it dotted black}) are from Porciani \& Madau (2001); the ``Fossil''
model ({\it solid red line}) is a simple broken power-law fit to
Nagamine et al.~(2006) (their Fig.~5); the ``GALEX'' model ({\it solid
blue line}) is the dust-corrected $\dot{\rho}_\star(z)$, adjusted to
the same assumed IMF as the other models presented here, from Figure 5
of Schiminovich et al.~(2005) ($\dot{\rho}_\star(z)$ assumed constant
for $z>3$); the ``H\&S'' model ({\it solid green line}) is an
analytic function from Hernquist \& Springel (2003).  The total TIR
background for these models as computed using eq.~(\ref{tir}) is
$F^{\rm TIR}=19.7$, 20.4, 23.0, 15.9, 19.5, and 17.3\,nW\,\,m$^{-2}$\,\,sr$^{-1}$
for RSF1, RSF2, RSF3, Fossil, GALEX, and H\&S models, respectively.  
These numbers for $F^{\rm TIR}$ can be compared with observational
constraints that indicate  $F^{\rm TIR}\sim24-27.5$\,nW\,\,m$^{-2}$\,\,sr$^{-1}$ (Dole et al.~2006).
Taking $f(z)=\min(0.9z+0.1,1)$, $f_{\rm cal}=0.79$, 0.81, 0.81 0.76, 0.75,
and 0.73 for these same models. Using these values of $F^{\rm TIR}$
and $f_{\rm cal}$, the total $\gamma$-ray and neutrino backgrounds can
be calculated from eq.~(\ref{gamma_neutrino_fir}) as in eq.~(\ref{background_total}). 
Figure \ref{fig:b} shows the cumulative fraction of $F^{\rm TIR}$ as a
function of redshift, $\zeta(z)$, for each of these models (eq.~\ref{zeta}).\\ 
\label{fig:sfr}}
\end{figure}
%%%%%%%%%%%%%%%%%%%%%%%%%%%%%%%%%%%%%%%%%%%%%%%%%%%%%%%%%%%%%%%%%%%%%%%%%%%%%

\section{The Starburst Background}
\label{section:background}

Because of the predicted one-to-one correspondence between the
neutrino, $\gamma$-ray, and radio emission of starbursts and their IR
emission, the diffuse background in $\gamma$-rays or neutrinos can be
normalized to the TIR background, $F^{\rm TIR}$.  Figure \ref{fig:sfr}
shows a number of models for $\dot{\rho}_\star(z)$ drawn from the
literature, based on observations of star-forming galaxies at
different redshifts.
The integrated background from star formation is then simply
\beq
F^{\rm TIR}=\frac{c}{4\pi H_0}\int_0^\infty\frac{\epsilon\dot{\rho}_\star
c^2}{(1+z)^2}\frac{dz}{E(z)},
\label{tir}
\eeq where $E(z)=[\Omega_m(1+z)^3+\Omega_\Lambda]^{1/2}$ and $\epsilon$ is an IMF-dependent
constant.  Results for
$F^{\rm TIR}$ for the various models shown in Figure \ref{fig:sfr} are
given in the caption and range from $F^{\rm TIR}_{20}\approx 0.8$ to
$F^{\rm TIR}_{20} \approx 1.15$, where $F^{\rm TIR}_{20}=F^{\rm
TIR}/20$\,nW\,\,m$^{-2}$\,\,sr$^{-1}$.  Note that our calculation here
does not include the order unity contribution to the optical/NIR
background at $z\approx0$ from the old stellar population (e.g., Nagamine et al.~2006; 
Dole et al.~2006).

\begin{figure}
\centerline{\hbox{\psfig{file=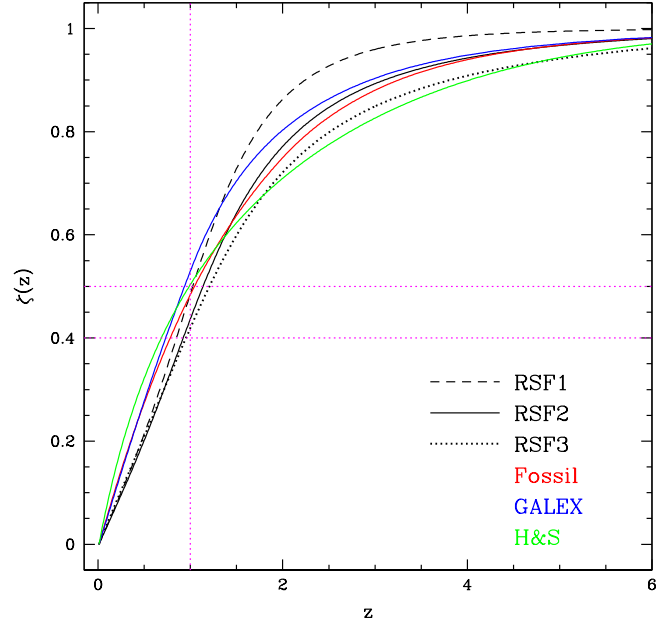,width=9cm}}}
\figcaption[x]{The cumulative contribution to the extra-galactic
background in starlight $\zeta(z)$ (see eq.~\ref{zeta}) as a function
of redshift for the models of $\dot{\rho}_\star(z)$ shown in Figure
\ref{fig:sfr}.  For reference, the horizontal magenta lines have been
put at $\zeta=0.4$ and $\zeta=0.5$, and the vertical line at $z = 1$.
Note that in all models roughly half of the extra-galactic background
comes from $z>1$. \\
\label{fig:b}}
\end{figure}

Figure \ref{fig:b} shows the contribution to $F^{\rm TIR}$ as a
function of redshift: \beq
\zeta(z)=\int_0^z\frac{\dot{\rho}_\star(z^\prime)dz^\prime}{(1+z^\prime)^2E(z^\prime)}/
\left\{\int_0^\infty\frac{\dot{\rho}_\star(z^\prime)dz^\prime}{(1+z^\prime)^2E(z^\prime)}\right\}.
\label{zeta}
\eeq The dotted lines indicate $z=1$ and $\zeta=0.4$ and 0.5, and have
been added for reference.  Note that all of the models consistent with
the observed star formation history of the universe have $\sim 1/2$ of
the TIR background produced at $z \gtrsim 1$.

%%%%%%%%%%%%%%%%%%%%%%%%%%%%%%%%%%%%%%%%%%%%%%%%%%%%%%%%%%%%%%%%%%%%%%%%%%%%
%%%%%%%%%%%%%%%%%%%%%%%%%%%%%%%%%%%%%%%%%%%%%%%%%%%%%%%%%%%%%%%%%%%%%%%%%%%%%

The total $\gamma$-ray and neutrino backgrounds can now be simply
estimated from the TIR background.  Equation
(\ref{gamma_neutrino_fir}) for the $\gamma$-ray and neutrino emission
from a given proton calorimeter implies that \beq
F_\gamma\approx(2/3)F_{\rm
nu}\approx1.4\times10^{-6}\eta_{0.05}f_{0.75}F^{\rm TIR}_{20}
\,\,\,\,{\rm GeV\,\,s^{-1}\,\,cm^{-2}\,\,sr^{-1}}
\label{background_total}
\eeq
where $f_{0.75}=f_{\rm cal}/0.75$ and
\beq
f_{\rm cal}=\int_0^\infty \frac{\dot{\rho}_\star(z)
f(z)\,dz}{[(1+z)^2E(z)]}/
\left\{\int_0^\infty
\frac{\dot{\rho}_\star(z)\,dz}{[(1+z)^2E(z)]}\right\}
\eeq
is the fraction of the TIR background produced by starburst
galaxies that are in the proton calorimeter limit (i.e.,
$\tau_{pp}<\tau_{\rm esc}$;
\S\ref{section:introduction}) and $f(z)$ is that
fraction at every $z$.  For a flat
$\gamma$-ray spectrum ($p\approx2$), equation (\ref{background_total})
implies a
specific intensity at GeV energies of
\beqa
\nu I_{\nu_\gamma} ({\rm GeV})&\approx& (2/3)\nu I_{\nu_{\rm nu}}
\nonumber \\
&\approx&
10^{-7}\eta_{0.05}f_{0.75}F^{\rm TIR}_{20} \,\,\,\,\,\,\,{\rm
GeV\,\,s^{-1}\,\,cm^{-2}\,\,sr^{-1}}.
\label{background}
\eeqa

Given the strong constraint on $\eta$ implied by the FIR-radio
correlation and the reasonable convergence of models and data on the
TIR background ($F^{\rm TIR}_{20}\approx1$; Fig.~\ref{fig:sfr}), the
estimated $\gamma$-ray and neutrino backgrounds depend primarily on
the fraction of star formation that occurs in the proton calorimeter
limit: $f(z)$ and $f_{\rm cal}$.  Locally, we estimate $f(z\approx0)
\sim 0.1$ from the fraction of the local FIR and radio luminosity
density produced by starbursts (e.g., Yun et al.~2001).  However, at
high redshift, a much larger fraction of star formation occurs in high
surface density systems that are likely to be proton calorimeters.
For example, the strong evolution of the IR luminosity function with
redshift implies that $f(z)$ increases dramatically from $z\approx0$ to
$z\approx1$ (e.g., Dole et al.~2006).
If we take $f(z\gtrsim1) \sim 1$ and $\dot{\rho}_\star(z)$ as in
any of the models of Figure \ref{fig:sfr},
we find that $f_{\rm cal} \sim 1$, with a significant contribution to the
$\gamma$-ray and neutrino backgrounds coming from $z \approx 1-2$.
In fact, we find that
\beq
F_\gamma\propto\nu I_\nu({\rm GeV})\propto f(z\approx1-2).
\label{highz}
\eeq More specifically, taking $\Omega_m=0.3$, $\Omega_\Lambda=0.7$, 
and assuming a function
$f(z)=\min(0.9z+0.1,1)$ that smoothly interpolates from a small local
starburst fraction ($f(z\approx0) \approx 0.1$) to an order unity
starburst fraction at high redshift, we find that $f_{\rm
cal}\approx0.8$ for the models shown in Figure \ref{fig:sfr} (see caption).  As an
example of another alternative, assuming $f(z)=\min[0.1(1+z)^3,0.8]$,
in which the starburst fraction increases roughly in proportion to
$\dot{\rho}_\star(z)$ and saturates at $f(z=1)=0.8$, $f_{\rm
cal}\approx0.6$ for the same models.  

For numerical calculations for the $\gamma$-ray background
specific to a given $f(z)$, $\dot{\rho}_\star(z)$,
and proton injection spectra, $p$, see Figure 1 of TQW.

\section{Comparison with Stecker (2006)}
\label{section:other}

In a recent paper, Stecker (2006) (S06v1, v2, v3) estimates a
significantly smaller $\gamma$-ray and neutrino background from
starburst galaxies than found in TQW and LW, respectively.  In his
original version of the paper (S06v1), Stecker normalized the
total neutrino background to the local FIR luminosity density from
starburst galaxies in Yun et al.~(2001) (based on {\it IRAS}), thereby
neglecting the contribution to the background from redshifts above
$z\approx0$.  This error led to a factor of $\approx$10 underestimate
of the total neutrino background relative to LW.

S06v2 corrects this error and addresses both the $\gamma$-ray and
neutrino backgrounds expected from starburst galaxies.  Stecker finds
a value for the neutrino and $\gamma$-ray backgrounds approximately
equal to the value quoted in S06v1, $\sim$5 times lower than LW and TQW.
There are two primary differences that lead to this discrepancy: \\

\noindent 1.~Stecker advocates that 22\% of the extra-galactic
background is produced by star formation above $z>1.2$ (see his Table
1, S06v3).  Figures \ref{fig:sfr} and \ref{fig:b} show, however, that
essentially all models for the star formation history of the universe
that are consistent with the observed evolution of the star formation
rate density with redshift ($\dot{\rho}_\star(z)$) yield roughly equal
contributions to $F^{\rm TIR}$ below and above $z\approx1$.  \\

\noindent 2.~We argue that order unity of all star formation at
$z\approx1$ occurs in starbursts likely to be proton calorimeters,
whereas Stecker quotes 13\% in the range $0.2\le z \le 1.2$ and 60\%
for $z>1.2$ (see his Table 1, S06v3). This step function model for
$f(z)$ likely significantly underestimates the contribution of 
$z \approx 1$ galaxies to the $\gamma$-ray and neutrino backgrounds. \\

Thus, the inconsistency between our predicted $\gamma$-ray and neutrino
backgrounds and Stecker's upper limit is a simple consequence of his
model for the star formation history of the universe and the redshift
evolution of the starburst population.  For the reasons explained
above, we believe that our assumptions are in better agreement
with observations of star-forming galaxies at $z \gtrsim 1$.

\section{Conclusion}
\label{section:conclusion}

The magnitude of the observed diffuse $\gamma$-ray background is 
quite uncertain, primarily as a result of foreground subtraction.
Whereas Sreekumar et al.~(1998) (compare with Strong et al.~2004)  
find a specific intensity of 
$1.4\times10^{-6}$\,GeV\,\,s$^{-1}$\,\,cm$^{-2}$\,\,sr$^{-1}$ at GeV energies, 
Keshet et al.~(2004) derive an {\it upper limit} of 
$\nu I_{\nu_\gamma}<5\times10^{-7}$\,GeV\,\,s$^{-1}$\,\,cm$^{-2}$\,\,sr$^{-1}$.
Future observations by {\it GLAST} should alleviate some of the 
uncertainty in the background determination.  A comparison of the 
upper limit by Keshet et al.~(2004) and equation (\ref{background})
indicates that starburst galaxies may indeed contribute significantly
to the diffuse $\gamma$-ray background.

Even in stacking searches, {\it EGRET} did not detect a single
starburst galaxy (Cillis et al.~2005).  In TQW (their Figure 1) we
show that the {\it EGRET} non-detections are required by predictions
that are correctly calibrated to the FIR or radio emission from
galaxies.  Hence, the {\it EGRET} non-detections do not provide a
significant constraint on the contribution of starburst galaxies to
the $\gamma$-ray or neutrino backgrounds.  {\it GLAST} will, however,
be much more sensitive.  Using equation (\ref{gamma_neutrino_fir}) and
numerical calculations for different $p$, TQW show that {\it GLAST}
should be able to detect several nearby starbursts (NGC\,253, M\,82,
IC\,342; see TQW Table 1) if they are indeed proton calorimeters.

The dominant uncertainty in assessing the starburst contribution to
the $\gamma$-ray and neutrino backgrounds lies in whether a significant
fraction of cosmic-ray protons in fact interact with gas of average density
or whether the cosmic rays escape in a galactic wind without coupling
to the high density ISM (the star formation history of the universe is
comparatively well understood).  Detection of a single starburst
galaxy by {\it GLAST} would strongly support the estimates of LW and
TQW and thus the importance of starburst galaxies for the
extra-galactic neutrino and $\gamma$-ray backgrounds.  Such a detection
would also provide an important constraint on the physics of the ISM
in starburst galaxies.

\end{document}